\pdfoutput=1
\documentclass[%
 reprint,
 amsmath,amssymb,
 aps,
floatfix,
]{revtex4-2}
\usepackage{amssymb}
\usepackage{amsmath}
\usepackage{xcolor}
\usepackage{hyperref}
\usepackage{cleveref}
\usepackage{amsthm}
\usepackage{enumitem}
\usepackage{graphicx}
\usepackage{dcolumn}
\usepackage{bm}

\newcommand{\blue}[1]{{\color{blue}#1}}

\begin{document}

\preprint{APS/123-QED}

\title{Multiple charge carrier species as a possible cause for triboelectric cycles}

\author{Juan Carlos Sobarzo}
 \email{jsobarzo@ist.ac.at}
\author{Scott Waitukaitis}%
 
\affiliation{%
 Institute of Science and Technology Austria, Am Campus 1, 3400 Klosterneuburg, Austria 
}%




\date{\today}

\begin{abstract}
The tendency of materials to order in triboelectric series has prompted suggestions that contact electrification might have a single, unified underlying description. However, the possibility of `triboelectric cycles,' \textit{i.e.}~series that loop back onto themselves, is seemingly at odds with such a coherent description.  In this work, we propose that if multiple charge carrying species are at play, both triboelectric series and cycles are possible.  We show how series arise naturally if only a single charge carrier species is involved and if the driving mechanism is approach toward thermodynamic equilibrium, and simultaneously, that cycles are forbidden under such conditions. Suspecting multiple carriers might relax the situation, we affirm this is the case by explicit construction of a cycle involving two carriers, and then extend this to show how more complex cycles emerge. Our work highlights the importance of series/cycles towards determining the underlying mechanism(s) and carrier(s) in contact electrification.

\end{abstract}

\maketitle


\section{\label{sec:intro}Introduction}

\begin{figure}[t!]
\centering
\includegraphics[width=0.45\textwidth]{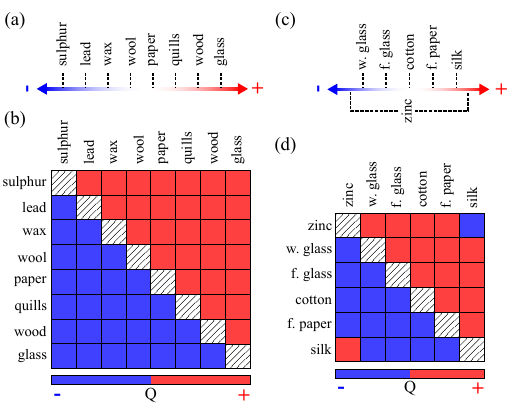}
\caption{ \label{fig:series_cycle} (a) The triboelectric series of Wilcke \cite{Assiss.2010}, where  materials are ordered by their tendency to charge plus/minus. (b) Sketch of what (a) may have looked like in matrix form. The color represents the sign of the charge transferred to the column material. (c) Shaw and Jex's triboelectric `cycle' \cite{Shaw.1928}.  Zinc charges positive to silk (at the top of the series) but negative to washed glass (at the bottom).  (d) Sketch of what (b) may have looked like in matrix form.}
\end{figure}

Contact electrification (CE)---the exchange of charge through contact---is among the oldest topics of scientific study \cite{Iversen.2012}, yet we understand little about it.  \blue{With metal-metal contacts, it is widely agreed that electrons are transferred, driven by the system trying to reach thermodynamic equilibrium where the Fermi surfaces are coincident \cite{Harper.1998, Lacks.2019}. This model is long-established, being proposed and initially validated by Harper as early as the 1950s \cite{Harper.1951}, and then convincingly re-validated by Lowell in the 1970s \cite{Lowell.1975}. Though few experiments are carried out with metals today, these tend to reconfirm and add to the existing framework \cite{Kaponig.2021}. With insulator-insulator contacts, the topic of this work, little is agreed upon: the identity of the charge carrier(s) is unknown, the mechanism(s) driving transfer are unresolved, and even whether or not it is an equilibrium process is a matter of debate \cite{Lacks.2019}.} There is a long history of experimental observations that suggest both `rhyme and reason' as well as puzzling inconsistency to insulator CE.
Among the most important of all observations is the tendency of materials to order in a triboelectric (TE) series. First observed by Johan Carl Wilcke in 1757, a TE series is an ordering of materials based on the sign of charge they acquire during contact \cite{Assiss.2010}. For example, Wilcke's series (Fig.~\ref{fig:series_cycle}(a)) has glass at the positive end, meaning it charged positive against other materials.  Wood charged negative to glass but positive otherwise, \textit{etc.}  When properly constructed, a series with $N$ materials requires $N(N-1)/2$ experiments, hence more information is present than a simple list can convey \cite{Zhang.20194bw}. More appropriately, one uses an $N$$\times$$N$ matrix, whose rows and columns indicate the materials involved and whose values indicate the sign of charge transferred (either to the column or row, depending on convention). For materials that order in a series, the rows/columns of the matrix can be rearranged such that all entries above (below) the diagonal have the same sign. For example, Fig.~\ref{fig:series_cycle}(b) shows what Wilcke's series would have looked like in matrix form, where the color indicates the sign of charge transferred to the column material. This representation will be important to our discussion.

Many investigations have sought to identify the material parameters that order TE series, 
\textit{e.g.}~acid/base properties \cite{Zhang.20194bw}, zeta potential \cite{McCarty.2008}, water adsorption properties \cite{Burgo.2016, Zhang.2015, Lee.2018, Waitukaitis.2014, Grosjean.2020, Grosjean.2023}, Seebeck coefficient \cite{Shin.2022}, cohesive energy density \cite{Sherrell:2021}, to name a few.  To date, it is fair to say that the CE community has not come to a consensus on which, if any, of such proposals is the correct one, and it is highly arguable that more than one might be at play \cite{Lacks.2019}.  What makes matters more complicated is the lore scattered throughout the literature regarding TE `cycles,' \textit{i.e.}~series that seem to `loop back onto themselves'. To our knowledge, the first mention of cycles comes from experiments conducted by Shaw and Jex in 1928 \cite{Shaw.1928}. One of their cycles---of which there were many---is shown in Fig.~\ref{fig:series_cycle}(c). Zinc charged negative to filter paper, cotton, fused glass and washed glass, whereas silk charged positive to these---yet zinc charged positive to silk.  When cast into matrix form (Fig.~\ref{fig:series_cycle}(d)), these results cannot be organized such that entries above/below the diagonal all have the same sign, no matter how the rows/columns in the matrix are ordered.  

Almost uniformly, subsequent references to TE cycles  \cite{ Zhang.20194bw, McCarty.2008, Shin.2022, Lacks.2012, Williams.2012, Soh.2012, Shaw.1928, Perez-Vaquero.2021, Pan.2019, Mutlu.2020, Lowell.1980,  Harper.1998, Gooding.2019, Dotterl.2016} merely point back to the results of Shaw and Jex, though some recent experiments measure cycles without drawing attention to them \cite{Sherrell:2021}. As observations of cycles are few, their existence is rightly a matter of debate, and in recent years the notion of TE series itself has been questioned. The traditional, perhaps optimistic view is that TE series are real and arise naturally due to thermodynamic considerations. This is appealing given the widespread observation of series, but incurs doubt due to discrepancies in series measured in different labs and by the notion of cycles. On the other hand, there is the view that both TE series and cycles are outdated notions, hopelessly compromised by uncontrollable influences from the nature of contact, surface roughness, mechanical stresses, humidity, \textit{etc.} \cite{Lacks.2019}---yet `throwing in the towel' doesn't explain why series are so often observed, nor does it explain or preclude cycles. Although these debates are not resolved, they highlight the centrality of TE series and cycles toward understanding the most important aspects of CE.

In this paper, we propose a simple `toy model' that allows for the existence of both TE series \textit{and} cycles, all within a framework explicitly based on equilibrium thermodynamics. 
Our essential idea that permits this is that more than one charge carrying species (\textit{e.g.}, electrons \textit{and} ions) is involved in CE. Given the charge carrier issue is a completely unresolved one \cite{Lacks.2019}, this is not at all unreasonable to consider.  We begin by assuming a single carrier, and show that TE series arise naturally if  these (\textit{i}) transfer between finite-depth energetic wells, and (\textit{ii}) approach toward thermodynamic equilibrium is the driving mechanism. We simultaneously show that, as often suspected but to our knowledge never demonstrated, the assumption of one carrier precludes TE cycles.  Next, we show by explicit construction that two charge carrier species can indeed permit the most basic TE cycle, \textit{i.e.}~with three materials. Finally, we extend our framework to account for more materials, considering in particular the different TE matrix configurations that may arise. Our work provides a rational framework for many widely held suspicions in the CE community, and highlights the significance of the existence of the TE series and cycles toward pinning down the underlying carrier(s) and mechanism(s).

\section{\label{sec:singlespecies}Equilibrium Model for a TE Series with a Single Charge Carrying Species}

\begin{figure}[t!]
\centering
\includegraphics[width=0.45\textwidth]{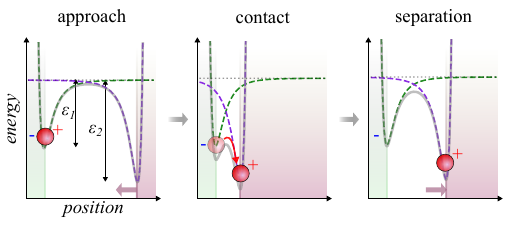}
\caption{ \label{fig:theoretical_framework} In our `toy model', we assume charge carriers inhabit energetic wells of depth, $\varepsilon$.  In the figure, the green curve corresponds to the well of the charge carrier on surface 1, the purple curve the well on surface 2, and the gray curve the sum of the two. When the two surfaces make contact, the wells `merge,' allowing the carrier to transfer between them.  At zero temperature, the carrier would always move toward the deeper well; however as we explain in the text, at non-zero temperature and when many carriers are involved, some fraction of carriers will move `uphill', and a thermodynamic picture is appropriate.}
\end{figure}

The model we pursue is based on robust and widely accepted experimental observations: in insulator CE, charge carriers are locally trapped before and after exchange. Before exchange, this is confirmed by the inability of insulators to conduct charge. After exchange, it is confirmed by Kelvin Probe Force Microscopy (KPFM) and Scanning Kelvin Probe (SKP) experiments, which reveal finite-size charge regions that are stable for hours or even longer, dependent on material and environmental conditions \cite{Heinert.2021, Bai.2021, Ji.2021, Burgo.2012, Sobolev.2022, Pertl.2022}.  Regardless of whether charge carriers are ions or electrons, these observations reveal they occupy energetic wells of finite depth on each surface (Fig.~\ref{fig:theoretical_framework}). This idea was put forth as early as 1957 by Henry \cite{Henry.1957}, and continues to receive interest \cite{McCarty.2008}. We assume charge transfer occurs as a result of carriers moving between wells as they merge/split during contact. Many other influences surely affect CE and could be considered (\textit{e.g.}~friction  \cite{Lowell.1980}, roughness effects \cite{Verners.2022}, surface water \cite{Burgo.2016, Zhang.2015}, mechanochemistry \cite{Mizzi.2022,Olson.2022,Fatti.2023}, \textit{etc.}), but are outside the scope of our work and do not affect our eventual conclusion.

If a charge carrier occupying a well on material $i$, with binding energy $\varepsilon_i$, is brought near to an empty well on material $j$, with binding energy $\varepsilon_j$, charge transfer may occur.  Macroscopic surfaces have many such well-pairs, which we must consider.  \blue{We denote the initial number of carriers on surfaces $i$ and $j$ as $n_i$ and $n_j$, respectively, which we call the `neutral numbers' since the surfaces are neutral with these numbers of carriers present. We consider the binding energies and neutral numbers `material properties', and furthermore we consider the neutral numbers as extensive. The rationale for this is that, whether charge carriers are electrons or ions, their interaction with a surface depends on the atoms present (including potential adsorbates), their relative proportions and structural arrangement---\textit{i.e.}~material properties. Due to the fact that properties of a material surface can change due to a myriad of factors, our definition of `material' is preparation specific. More generally, one could consider that for identically prepared samples of a single material, the neutral numbers could be drawn from a distribution inherent to the material. This would allow, for example, ``same-material'' tribocharging. For simplicity, however, we do not delve into this possibility.} During contact, some pairs of merging wells may have transfer $i\rightarrow j$, others $j\rightarrow i$, and others not at all. Assuming the final distribution of charge is caused by the system moving toward equilibrium, we can determine the final configuration. 


\blue{We therefore look at the partition function of the system.  We make the assumption that the surfaces come into normal, frictionless contact and that each charge carrier is independent from all others. The situation is therefore that of a two-well system in a heat bath with a constant temperature $T$, and can be treated with the canonical ensemble.  Strictly speaking, the partition function is $Z = e^{\varepsilon_i / \kappa T} + e^{(\varepsilon_j+\Delta U(d) )/\kappa T} $, with $\kappa$ the Boltzmann constant.} The term, $\Delta U(d)$, reflects the fact that if the surfaces have net charge due to the transfer, a long range electrostatic energy is present that grows with the separation, $d$ \cite{Henry.1957}. Such long-range interactions are also considered in the metal-metal case \cite{Harper.1998} and lead to `back-tunneling' of charge after contact, but are cutoff at very small length scales ($\sim$1 nm) due to the increasing difficulty of tunneling with separation. Cutoff lengths should be even shorter in other circumstances (\textit{e.g.}~for ions \textit{vs.}~electrons). We therefore approximate this term as null, and the partition function simplifies to $Z = e^{\varepsilon_i/\kappa T}+e^{\varepsilon_j/\kappa T}$.  Thus, the average number of carriers on surface $j$ in equilibrium is
\begin{equation}
\color{blue}
 \bar{n}_{ij} = \frac{(n_i+n_j)e^{\varepsilon_j/\kappa T}}{e^{\varepsilon_i/\kappa T}+e^{\varepsilon_j/\kappa T}},
\label{eq:avg_numbers}
\end{equation}
and, naturally, for surface $i$, $\bar{n}_{ji}=n_i+n_j-\bar{n}_{ij}$ due to charge conservation. Taking for now that the charge carriers are positive and unity, and assuming the surfaces start out neutral before contact and approach equilibrium after, the total charge given from surface $i$ to $j$ is $q_{ij} \equiv \bar{n}_{ij} - n_j$ (where the first index transfers charge to the second). Note as well that charge conservation implies $q_{ji}=-q_{ij}$. Surface $j$ charges positively when $q_{ij}>0$, which given Eq.~\ref{eq:avg_numbers} leads to the condition
\begin{align}
\frac{n_j}{n_i} < e^{(\varepsilon_j - \varepsilon_i )/\kappa T} .
\label{eq:condition12}
\end{align}
If instead we had assumed negative charge carriers, the inequality would be flipped. This result highlights that the sign of charge a surface acquires is not driven solely by energetic differences, but also by differences in neutral numbers \cite{Henry.1957}. For example, if $\varepsilon_i = \varepsilon_j$ the condition reads $n_j<n_i$, which is equivalent to saying that carriers diffuse from regions of higher to lower concentration.

Now, we extend this single-carrier framework to show that it (\textit{i}) leads naturally to TE series, and (\textit{ii}) simultaneously precludes cycles. We consider three materials, $i$, $j$, and $k$, and assume that $j$ charges positively to $i$, and that $k$ charges positively to $j$, \textit{i.e.}
\begin{align}
\frac{n_j}{n_i} & < e^{(\varepsilon_j - \varepsilon_i )/\kappa T}  \\
\frac{n_k}{n_j} & < e^{(\varepsilon_k - \varepsilon_j )/\kappa T}.
\end{align}
Multiplying these two equations together, we have
\begin{equation}
\frac{n_k}{n_i} < e^{(\varepsilon_k - \varepsilon_i )/\kappa T},
\label{eq:series_proof}
\end{equation}
which is equivalent to the statement that $k$ charges positively to $i$.  Hence, this establishes that our three materials with a single charge carrier species will order into a series. Since a cycle would require replacing the less-than sign of Eq.~\ref{eq:series_proof} with a greater-than sign, it simultaneously shows cycles are not possible.  Conveniently, using indices such as `$i$' and `$j$' naturally translates to matrix notation when an arbitrary number of materials, $N$, are involved.  As mentioned previously, this matrix has the property $q_{ij} = -q_{ji}$, and if the materials order into a series it can have the rows/columns swapped such that all entries above/below the diagonal are the same sign, as in Fig.~\ref{fig:series_cycle}(b).

\section{\label{sec:2species}Explicit Construction of a Cycle with Two Charge Carrying Species}

\begin{figure}[h!]
\centering
\includegraphics[width=0.45\textwidth]{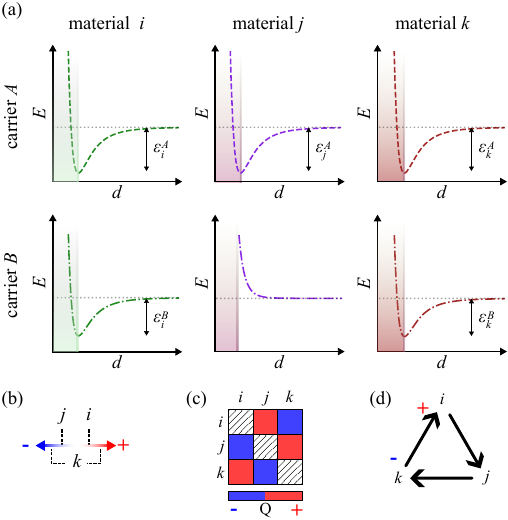}
\caption{ \label{fig:two_carriers} (a) We explicitly construct a cycle with two carriers and three materials.  Assuming positive carriers $A$ and $B$, $A$ alone would give the series $\{i,j,k\}$, while $B$ alone would give $\{k,i\}$.  (b)  In combination, and with appropriately chosen neutral numbers, a 3-cycle $\langle i,j,k\rangle$ is possible.  (c) Matrix representation of the $3$-cycle. (d) Arrow diagram to visualize the relations between surfaces when cycles are involved. An arrowhead pointing from $k$ to $i$ implies that charge transfers in that direction, or equivalently, $k$ charges negative and $i$ charges positive.
}
\end{figure}

We now show with a particular example that cycles are possible when more than one carrier is present, assuming they transfer independently. We consider three materials, $i,j,k$, and two carriers, $A$ and $B$. For carrier $A$, we assume the particular values $\varepsilon_i^A=\varepsilon_j^A=\varepsilon_k^A$. For carrier $B$, we assume $\varepsilon_i^B=\varepsilon_k^B$, but for material $j$ there is no well---\textit{i.e.}~surface $j$ has no sites to accommodate carrier $B$, or effectively $\varepsilon_j^B=-\infty$ (see Fig.~\ref{fig:two_carriers}(a)). This choice for $\varepsilon_j^B$ is not a necessary one, but it is a particularly convenient one as it allows us to construct a cycle that is obvious and requires little math. Regarding neutral numbers, we assume for carrier $A$ that $n_i^A>n_j^A>n_k^A=0$, and for carrier $B$ that $n_k^B>n_i^B=n_j^B=0$. Under these assumptions, transfer of carrier $A$ alone would give the series from negative to positive, $\{i, j, k \}$, while transfer of $B$ alone would give the series $\{ k, i \}$---material $j$ excluded since no transfer of $B$ occurs when $j$ is involved.  When both carriers transfer simultaneously, the total charge is now the sum of the two,
\begin{align}
q_{i j} & = q_{i j}^A \label{eq:charge_1}, \\
q_{j k} & = q_{j  k}^A \label{eq:charge_2}, \\
q_{k i} & = q_{k  i}^A + q_{k i}^B .\label{eq:charge_3}
\end{align}

One particular set of conditions that define a cycle is when $q_{i  j},q_{j  k}, q_{k  i} > 0$. Recalling Eq.~\ref{eq:condition12}, this leads to the inequalities,
\begin{align}
n_j^A & < n_i^A, \label{eq:inequality_1}\\
0 & < n_j^A, \label{eq:inequality_2}\\
n_i^A & < n_k^B. \label{eq:inequality_3}
\end{align}
Equations \eqref{eq:inequality_1} and \eqref{eq:inequality_2} are trivially satisfied, as they simply reiterate the assumed neutral numbers for carrier $A$ to produce the series $\{i,j,k\}$. On the other hand, Eq.~\eqref{eq:inequality_3} merely requires that the neutral number of carrier $B$ on $k$ is greater than that of carrier $A$ on $i$, which is not at all at odds with our assumptions. Hence, within this equilibrium picture, the existence of two charge carrier species can produce a cycle (Fig.~\ref{fig:two_carriers}(b)). Cast into matrix form, the cycle we have constructed here is as shown in Fig.~\ref{fig:two_carriers}(c), where the entries above (below) the diagonal cannot be arranged to have a single sign, regardless of any switching of rows/columns. As a conceptual aid, we show in Fig.~\ref{fig:two_carriers}(d) the same information in the form of an arrow diagram, which indicates the flow of positive charge.

\section{\label{sec:complexcycles}Complex Cycles of Higher Order}

\begin{figure*}
    \centering    \includegraphics[width=\textwidth]{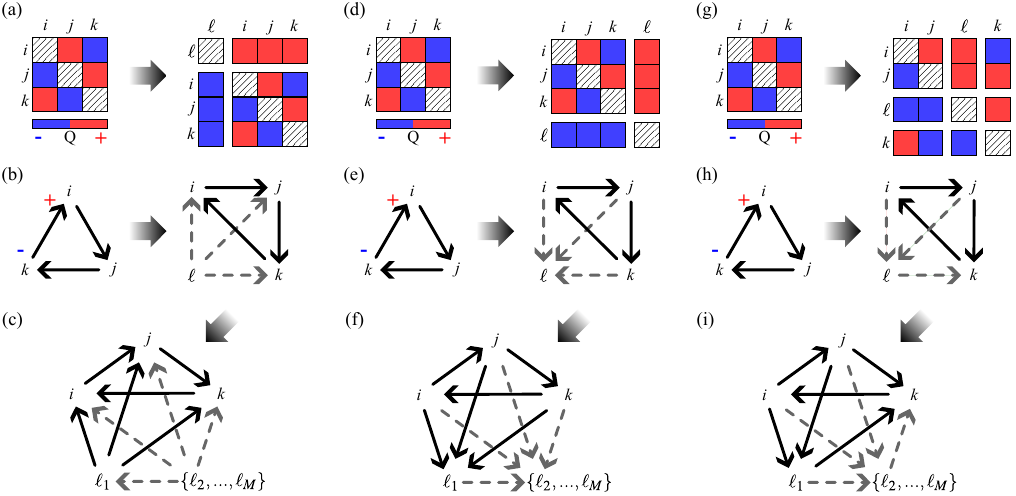}
    \caption{Creation of more complex cycles through insertion of new materials (a-c) before the existing cycle, (d-f) after, and (g-i) in-between materials. Figures (a),(d), \& (g) show the matrix representation of the material insertion process for a $3$-cycle, while the respective arrow diagrams (b),(e) \& (h) provide visual aid to understand the relation between surfaces in each case. The process can be extended to introduce an arbitrary number of new materials, as shown in diagrams (c), (f), \& (i). }
    \label{fig:complex_cycles}
\end{figure*}

We now generalize to more complex situations. We define the \textit{order} of a cycle as the number of materials that comprise it---an $n$-cycle is formed by $n$ materials; \textit{e.g.}~the cycle of the previous section is a $3$-cycle. We now show that materials can be added to existing cycles in different ways. We presume the existence of a $3$-cycle with materials $i$,$j$ and $k$, such that $q_{i j}, q_{j k}, q_{k i} >0$. We now consider a new material, $\ell$, and ask whether or not it can have ($n_\ell^A, n_\ell^B, \varepsilon_\ell^A, \varepsilon_\ell^B$) so it creates specific charging configurations. In principle, there are eight possibilities, since $\ell$ can charge two ways (positively or negatively) with each of the three other materials. Incidentally, the charging matrices of several of these are isomorphic, so only three need be investigated.

The first configuration is that $\ell$ charges negative to $i$, $j$ and $k$. The matrix for this is shown in Fig.~\ref{fig:complex_cycles}(a), where we have the `insertion' of $\ell$ before $i$. We now take a perturbative approach, and assume all parameters of $\ell$ are identical to $i$ except for $n_\ell^A$, which is infinitessimally different from $n_i^A$, \textit{i.e.}~
\begin{equation}
n_\ell^A = n_i^A + \eta.
\label{eq:def_n_eps}
\end{equation}
Following this, we find the requirement that $\eta$ must satisfy
\begin{equation}
0 < \frac{q_{k i}}{f(\varepsilon_i^A,\varepsilon_k^A)} < \eta , \label{eq:eta_1}
\end{equation}
where
\begin{equation}
\color{blue}
f(\varepsilon_i^A,\varepsilon_k^A)\equiv \frac{e^{\varepsilon_k/\kappa T}}{e^{\varepsilon_i/\kappa T}+e^{\varepsilon_k/\kappa T}} .\label{eq:def_f}
\end{equation}
Hence, if a 3-cycle exists, it is possible to `add a material before it' that charges negatively to all materials in the cycle. As a visual aid, Fig.~\ref{fig:complex_cycles}(b) shows an arrow diagram that illustrates the relation between surfaces after the insertion of the additional material. Note that this configuration leaves the existing $3$-cycle unaffected and there are no new cycles created as a consequence. 

From the previous analysis, it is straightforward to show that an arbitrary number of materials $M$ can be introduced consecutively, provided that the neutral numbers for $A$ are defined incrementally, \textit{i.e.} $n_{\ell_m}^A = n_i^A + m \eta$, where $m$ is an index on the set of new materials $\ell_m=\{\ell_1,\ell_2,...,\ell_M\}$. The rest of parameters are identical to those of $i$.  Then $\eta$ must satisfy
\begin{equation}
0 < \frac{q_{k i}}{M f(\varepsilon_i^A,\varepsilon_k^A)} < \eta .\label{eq:eta_M_1}
\end{equation}
The extension of the process again leaves the original cycle untouched, as seen in Fig.~\ref{fig:complex_cycles}(c). Naturally, this analysis for the addition of materials `before' a cycle can be extended to the addition of materials `after' one as well, as illustrated in Fig.~\ref{fig:complex_cycles}(d-f). Hence cycles can be `embedded' in arbitrarily large series.


The cases considered so far leave the original cycle untouched. However, it is also possible to add new materials in such a way that new and more complex cycles are created. We again start with the same $3$-cycle, but now the charge exchanges for the new surface $\ell$ must satisfy $q_{\ell i}, q_{\ell j} < 0$ and
$q_{\ell k} > 0$, that is, we put $\ell$ between $j$ and $k$ (Fig.~\ref{fig:complex_cycles}(g)). Under the same assumptions for the parameters as for the previous case, this time $\eta$ must satisfy
\begin{equation}
-\frac{q_{j k}}{f(\varepsilon_j^A,\varepsilon_k^A)} < \eta < 0 . \label{eq:eta_3}
\end{equation}
In Fig.~\ref{fig:complex_cycles}(h) we observe that the insertion of a new material in this manner has created an additional $3$-cycle: $\langle i,\ell,k\rangle$, as well as a $4$-cycle: $\langle i,j,\ell,k\rangle$. We remark that this case is isomorphic to all five of the remaining configurations for the insertion for $\ell$.  

Furthermore, if we add an arbitrary number of materials $M$ by a process analogous to the previous cases, we obtain the following conditions for the parameter $\eta$:
\begin{equation}
- \frac{q_{j k}}{M f(\varepsilon_j^A,\varepsilon_k^A)} < \eta < 0 . \label{eq:eta_M_3}
\end{equation}
Thus, we can create a cycle of arbitrary order $M$, as shown in Fig.~\ref{fig:complex_cycles}(i), and hence recover the Shaw and Jex-type $6$-cycle in Fig.~\ref{fig:series_cycle}(c,d). Additionally, such configuration has the consequence of creating new $3$-cycles with each new material added: $\langle i,\ell_m, k\rangle$, as well as $4$-cycles with pairs of these new materials, \textit{e.g.}, $\langle i,\ell_1,\ell_2,k\rangle$, and potentially, a number of cycles of increasing orders up to $M$.

This line of analysis shows that it is possible not only to reproduce a Shaw and Jex's type of cycle, but also to create other types with a variety of orders and configurations.  It also highlights the fact that inconsistencies observed in a reported TE series when put in matrix representation actually translate to cycles hiding in plain sight, such as the one in reference \cite{Sherrell:2021}.

\section{\label{sec:discandconc}Discussion and Conclusions}

\blue{The situation we have considered is straightforward: charge carriers are assumed to live in and transfer between energetic wells, the number of carriers and well depths are material parameters, and the final distribution of carriers is set by thermodynamic equilibrium.  In this context, we showed that a single charge carrier species implies only TE series, while more than one allows for the possibility of TE cycles.  We further showed that for more than three materials, more complex situations are possible.  For example, materials can be added to a cycle in such a way that new cycles are created or avoided.}

\blue{Our focus on a `toy model' leaves out many important considerations that are pertinent to insulator CE.  First, we have assumed that the physics underlying charge transfer is driven by the approach to equilibrium. The field fails to agree if this is the case or not \cite{Lacks.2019}.  Conceptually, the equilibrium we consider can only be established over surface regions that are close enough to be considered in contact, since charge carriers in other regions cannot move.  Our argument that neutral numbers are extensive material parameters means that the sign of charging depends only on the materials, not inevitable changes in the contact area---hence, within our assumptions, our conclusions about series/cycles survive this subtlety.  Moreover, we have assumed that the contact is normal and frictionless, hence does not involve the non-equilibrium ingredients of localized heating or stress.  This is motivated by the fact that even the most careful experiments meant to avoid rubbing/sliding still observe charge exchange, and by the fact that `hot spot' models for charging have not been widely accepted as a (let alone `the') cause for CE \cite{Lacks.2019}.  Lastly, the idea that the neutral numbers are material properties can itself be challenged, leading to the possibility for other causes for cycles.  For example, if neutral numbers are allowed to vary significantly from one sample of the same material to the next, then even a single carrier could lead to the appearance of cycles.  The `rhyme and reason' found in CE, and specifically the prevalence of series, suggests this is not the case.}

Naturally, a model like ours asks for clever and careful experiments to test its validity. This is unfortunately not possible given the current state of the field---there is no agreement on identifying any \textit{one} charge carrier species involved in CE, let alone on the possibility that two or more could be at play. Given this reality, the primary benefit of our work is not to predict the ordering of a particular TE series in different experiments or whether or not a specific cycle would occur. \blue{Instead, the value of this work is to: (a) present a rational explanation for how cycles can exist, (b) show that the simultaneous existence of series and cycles need not be contradictory, and (c) show that cycles do not require a non-equilibrium process. These insights highlight how important series/cycles are toward understanding the fundamental carrier(s) and mechanism(s) of CE. In other words, we make an effort to address the 'elephant in the room' of CE, as TE series and cycles are so often mentioned in the literature but they are rarely discussed in depth, neither theoretically nor experimentally.}

\begin{acknowledgments}
We would like to thank Carl Goodrich for helpful discussions. This project has received funding from the European Research Council (ERC) under the European Union’s Horizon 2020 research and innovation programme (Grant agreement No.~949120).
\end{acknowledgments}


\bibliographystyle{apsrev4-2}
\bibliography{apssamp}

\end{document}